\documentclass{article}

\usepackage{arxiv}
\usepackage{cite}
\usepackage{amsmath,amssymb}

\usepackage[utf8]{inputenc} %
\usepackage[T1]{fontenc}    %
\usepackage{hyperref}       %
\usepackage{url}            %
\usepackage{booktabs}       %
\usepackage{amsfonts}       %
\usepackage{nicefrac}       %
\usepackage{lipsum}
\usepackage{graphicx}
\graphicspath{{./figures/}{../figures}}

\title{Topological Measures for Pattern quantification of Impact Centers in Piezo Vibration Striking Treatment (PVST)}

\author{
 Melih C. Yesilli\\
    Mechanical Engineering\\
	Michigan State University\\
	East Lansing, MI 48824\\
    Email: yesillim@egr.msu.edu\\
    (Corresponding Author)
   \And
Max M. Chumley\\
    Mechanical Engineering\\
	Michigan State University\\
	East Lansing, MI 48824\\
    Email: chumleym@msu.edu
  \And
 Jisheng Chen \\
    Mechanical Engineering\\
	Michigan State University\\
	East Lansing, MI 48824\\
    Email: chenjish@msu.edu
    \And
 Firas A.~Khasawneh\\
    Mechanical Engineering\\
	Michigan State University\\
	East Lansing, MI 48824\\
    Email: khasawn3@egr.msu.edu
    \And
 Yang Guo\\
    Mechanical Engineering\\
	Michigan State University\\
	East Lansing, MI 48824\\
    Email: yguo@msu.edu\\
}

\begin{document}
\maketitle
\begin{abstract}
Surface texture influences wear and tribological properties of manufactured parts, and it plays a critical role in end-user products. Therefore, quantifying the order or structure of a manufactured surface provides important information on the quality and life expectancy of the product. Although texture can be intentionally introduced to enhance aesthetics or to satisfy a design function, sometimes it is an inevitable byproduct of surface treatment processes such as Piezo Vibration Striking Treatment (PVST). Measures of order for surfaces have been characterized using statistical, spectral, and geometric approaches. For nearly hexagonal lattices, topological tools have also been used to measure the surface order. This paper utilizes tools from Topological Data Analysis for quantifying the impact centers' pattern in PVST. We compute measures of order based on optical digital microscope images of surfaces treated using PVST. These measures are applied to the grid obtained from estimating the centers of tool impacts, and they quantify the grid's deviations from the nominal one. Our results show that TDA provides a convenient framework for the characterization of pattern type that bypasses some limitations of existing tools such as difficult manual processing of the data and the need for an expert user to analyze and interpret the surface images. 
\end{abstract}

\section{Introduction}
\label{sec:Intro}

One of the main objectives of the manufacturing enterprise is to achieve products that satisfy a preset quality under the constraints of time, cost, and available machines \cite{Benardos2003}. 
A key quality in manufacturing is the surface texture which is directly related to surface roughness \cite{Iso4287,Iso25178,ASMEB46p1} and to the tactile feel of the resulting products which is quantified by tactile roughness \cite{Dahiya2010,Ding2017a}. 
Surface texture can be either intentionally introduced to satisfy functional or aesthetic surface properties, or it can be a byproduct of a specific manufacturing setting. 
However, the importance of surface texture goes beyond merely the aesthetics since the resulting surface properties have a strong influence on ease of assembly, wear, lubrication, corrosion \cite{Thomas1998}, and fatigue resistance \cite{Spierings2013,Frazier2014,Chan2012}. 

\begin{figure*}[htbp]
\centering
\includegraphics[width=0.95\textwidth,height=1\textheight,keepaspectratio]{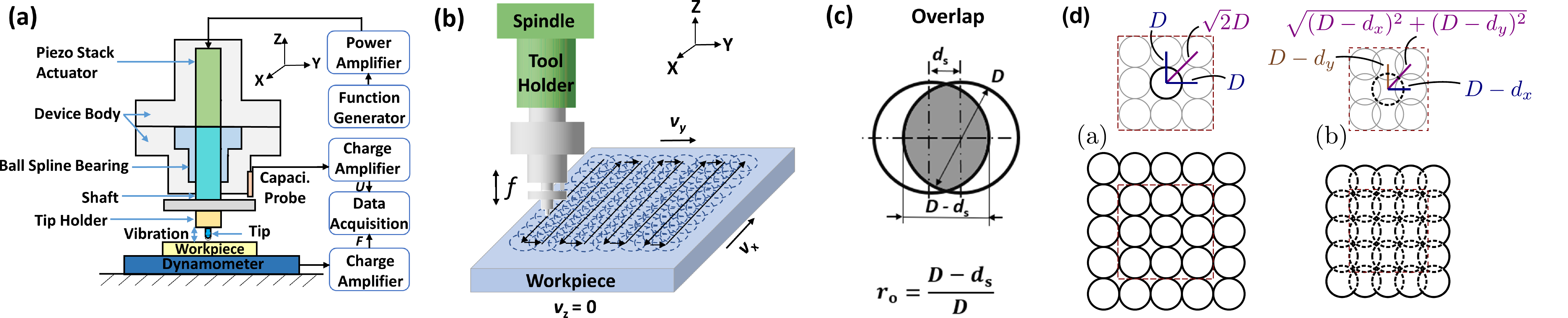}
\caption{Schematics of PVST: (a) PVST; (b) Striking process in PVST; (c) Overlap of indentations; (d) Grid geometry. 
PVST nominal grid for (d1) $x$ and $y$ overlap $d_x=d_y=0$, and d2) for $x$ overlap $d_x \geq 0$ and $y$ overlap $d_y\geq 0$. The striking tool diameter is $D$, and the figure shows a scanning strategy where the $x$ and $y$ scanning speeds are set to obtain a certain overlap ratio. When the overlap in $x$ and $y$ is the same, we write $d_x=d_y=d_s$.
}
\label{fig:PVST}
\end{figure*}

An example of a process where surface texture is introduced in order to both enhance the mechanical properties of the part and as an inevitable byproduct is the Piezo Vibration Striking Treatment (PVST) \cite{chen2021}. 
In PVST, a tool is used to impact the surface and a scanning strategy is applied to treat the whole surface, see Fig.~\ref{fig:PVST}. Depending on the impact depth and speeds set for the process, various grid sizes and diameters can be obtained. %
By varying parameters such as the scanning speed and the overlap of the impacts, a texture inevitably is left behind on the surface. 
While this can be leveraged to both treat and texture the surface, the resulting pattern can also provide invaluable information about the success of the treatment such as quantifying missed or misplaced impact events. 
Further, the texture can also be utilized to assess the quality of the machined surface through comparing the resulting pattern with the nominal or desired pattern. 
Specifically, if the resulting pattern is missing too many features (indentations), this can be an indication of large deviation of material distribution on the surface. 
Detecting such events can signal the need for further finishing, or for adjusting the manufacturing process to enhance the resulting surfaces. 

While there are several classical tools for quantifying surface texture \cite{Hossain2012}, one limitation of these methods is their strong reliance on the user for tuning the needed parameters. For example, a common pre-requisite for these tools is knowing the relative pixel intensity in the image and which threshold size for removing objects from the image will result in a successful texture segmentation.  
An emerging tool that has shown promise for quantifying texture is from the field of Topological Data Analysis (TDA). 
More specifically, topological measures were used to quantify order of nearly hexagonal lattices \cite{Motta2018} which represent nanoscale pattern formation on a solid surface that can result from broad ion beam erosion \cite{Facsko1999}. 
The input data in \cite{Motta2018} was the location of the nanodots on the surface, which is an input data type often referred to as a point cloud. 
Topological measures were shown to be more sensitive than traditional tools, and they can provide insight into the generating manufacturing process by examining the resulting surfaces. 

This paper explores utilizing topological measures for quantifying the surface texture produced by PVST. 
Specifically, the goal of this paper is to use topological methods to quantify lattice types in a PVST image, which can provide insight into the effectiveness of the PVST process. 
While we use measures inspired by those utilized on point clouds, i.e., point locations in the plane, in \cite{Motta2018}, the data in our work are images of the resulting surface obtained using KEYENCE Digital  Microscope.  
Therefore, in our setting we need to process the data to extract the PVST indentation centers in order to quantify the resulting pattern. 
Our exploratory results show possible advantages to our approach including automation potential in contrast to standard tools where intensive user-input is required. 

The paper is organized as follows. Section~\ref{sec:PVST} provides background for PVST. Section~\ref{sec:Experimental Setup} explains the experimental setup and how the experimental data is collected. Section \ref{sec:data_processing} outlines the processing performed on each image to obtain the point cloud data. Section~\ref{sec:TDA} discusses the TDA-based approach proposed in this study. Section~\ref{sec:Results} compares the results of the analysis to a perfect square lattice. Section~\ref{sec:Conclusion} includes the concluding remarks.

\section{Piezo Vibration Striking Treatment}
\label{sec:PVST}
Mechanical surface treatment uses plastic deformation to improve surface attributes of metal components, such as surface finish, hardness, and residual stress, which is an effective and economical way of enhancing the mechanical properties of engineering components. Among various mechanical surface treatment processes, i.e., Shot Peening \cite{Torres2002}, Surface Mechanical Attrition Treatment \cite{Roland2006}, High-frequency Mechanical Impact Treatment \cite{Yildirim2012}, and Ultrasonic Nanocrystal Surface Modification \cite{Cao2010}, PVST is a novel mechanical surface treatment process that is realized by a piezo stack actuated vibration device integrated onto a computer numerical control (CNC) machine to impose tool strikes on the surface. Different from those processes, the non-resonant mode piezo vibration in PVST and the integration with CNC machine enable PVST to control the process more conveniently and precisely as demonstrated in previous applications in modulation-assisted turning and drilling processes \cite{Guo2017,Guo2020}. 
The schematics of PVST are shown in Fig.~\ref{fig:PVST}. The device is connected to the spindle of a CNC mill through a tool holder. The spindle can only move along the $Z$ direction to control the distance between striking tool and workpiece surface. The motion of the machine table along the $X$ or $Y$ directions defines specific striking locations on the workpiece mounted on the table. As shown in Fig.~\ref{fig:PVST}a, the piezo stack actuator is connected with a spline shaft that is part of the ball spline bearing, both of which are fixed in the device body. The actuator drives the shaft to move along the $Z$ direction, but no bending or rotation is allowed. The striking tool is rigidly connected to the shaft through a holder. The power generator and amplifier produce amplified driving voltage to extend and contract the actuator and hence actuate the tool to oscillate along the axial direction. A capacitance probe clamped onto the device body and a dynamometer plate mounted on the machine table are used to measure the displacement of the tool and the force during the treatment. Both the force and displacement are recorded synchronously in a data acquisition system. The lower bound and upper bound of the driving voltage are set as zero and peak-to-peak amplitude $V_{pp}$ of the voltage oscillation, which can control the frequency and amplitude of the tool vibration to generate different surface textures. The initial position of the tool $Z$ can be used to control the distance between the tool and the workpiece surface and hence change the striking depth to produce different surface textures as well. As shown in Fig.~\ref{fig:PVST}b-d, the successive strikes controlled by scan speed $v_{s}$ will be imposed on different locations of the surface along the tool scan path. The offset distance $d_{s}$ between two successive strikes and the diameter of the indentation can be utilized to compute the overlap ratio. Different overlap ratios can generate various surface textures, namely higher overlap ratio leads to a denser distribution of the indentations. This paper focuses on the low overlap ratio images produced using PVST to detect the center points of the circles in the image and quantitatively determine the lattice type present in the texture with minimal user input.  

\begin{figure}[htbp]
\centering
\includegraphics[width=0.45\textwidth,height=1\textheight,keepaspectratio]{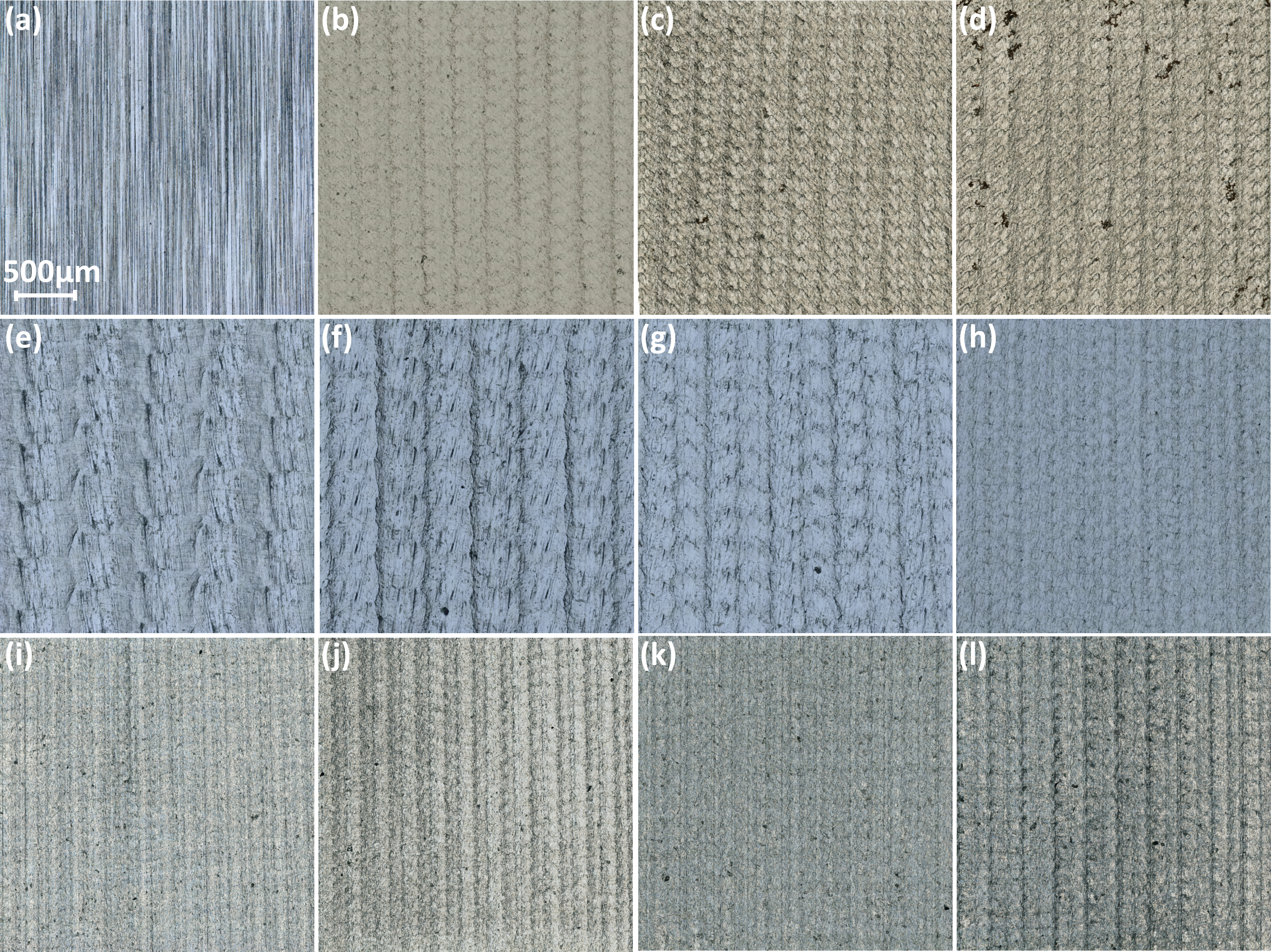}
\caption{Various surface textures under different PVST conditions: (a) initial workpiece surface; (b) – (d) different Z values; (e) – (h) different overlap ratios; (i) – (l) different driving voltages.
}
\label{fig:samples}
\end{figure}

\section{Experimental Procedure}
\label{sec:Experimental Setup}
A mild steel ASTM A572GR50 workpiece with a dimension of 120 mm × 40 mm × 20 mm is used for surface texture data collection under various PVST conditions (see Tab.~\ref{tab:PVST}). 

\begin{figure}[htbp]
\centering
\includegraphics[width=0.45\textwidth,height=1\textheight,keepaspectratio]{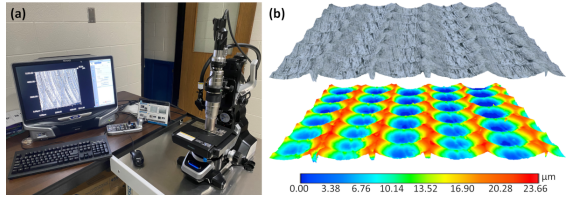}
\caption{(a) KEYENCE digital microscope (b) illustrations of scanned surface texture.}
\label{fig:device}
\end{figure}

\begin{table}[htbp]
\centering
\caption{Various PVST conditions. The first column represents the samples in Fig.~\ref{fig:samples}.}
\label{tab:PVST}
\begin{tabular}{cccccc}
\multicolumn{1}{c|}{No.}& f (Hz) & $V_{pp}$ (V)& d (mm) & $r_{o}$ & Z ($\mu$m)\\
\hline
b-d & 100 & 120 & 3 & 0.75 & 0,10,20\\
e-h & 100 & 120 & 3 & 0, 0.25, & 0\\
    &     &     &   & 0.5, 0.75 &  \\
i-l & 100 & 60, 90, & 3 & 0.75 & 0\\
    &     & 120, 150         &   &      &  \\
\hline
\end{tabular}
\end{table}
The treated area for each condition is 5 mm × 5 mm. Only a size of 2.5 mm × 2.5 mm is used for surface texture data collection due to the duplicate characteristic of the surface texture throughout the treated area and the computation efficiency for data processing. The selected area is fixed at the upper left corner of the treated area for consistent data collection. 
The workpiece is placed on a free-angle $XYZ$ motorized observation system (VHX-S650E), and 3D surface profiles are characterized using KEYENCE Digital Microscope (VHX6000),  as shown in Fig.~\ref{fig:device}a. A real zoom lens (KEYENCE VH-Z500R, RZ x500 - x5000) and x1000 magnification are utilized to achieve sufficient spatial resolution (0.21 $\mu$m). Since each capture under this magnification can only cover a small area, the stitching technique (22 × 22 scans in horizontal and vertical directions) is employed to achieve sufficient capture field. Fig.~\ref{fig:device}b shows one of the captured 3D profiles under two different types of illustrations (texture and surface height map). The scanned surface textures after different PVST conditions are shown in Fig.~\ref{fig:samples}. Note that for this paper, images \emph{e}, \emph{f}, and \emph{g} were the main focus because they allow extracting the indentation centers. The centers in the remaining images are not easily identifiable, and they require tools from image analysis that are beyond the scope of this paper.

\section{Data Preprocessing}\label{sec:data_processing}
\begin{figure}[htbp]
\centering
\includegraphics[width=0.45\textwidth,height=1\textheight,keepaspectratio]{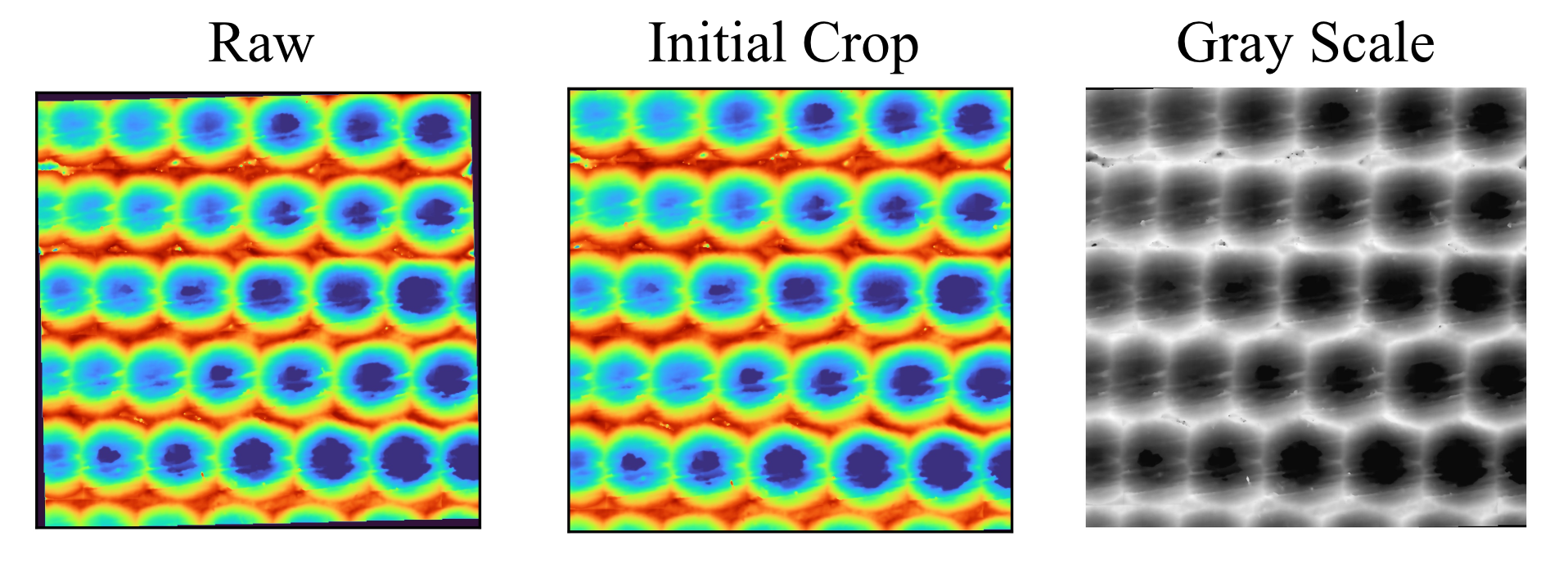}
\caption{Preprocessing of a sample raw image. }
\label{fig:preprocess_120_0}
\end{figure}

In this section, we explain how we preprocess the data set. The raw images obtained from the microscope have dimension of nearly $12000\times 12000$.
However, there are black pixels around the edges of each image. 
We removed these pixels such that all raw images have a final dimension of $12000\times 12000$ taken from the center of the corresponding raw image. 
This significantly reduced the number of black pixels in the retained images.
Each image was then converted to grayscale as shown in Fig.~\ref{fig:preprocess_120_0}. 

\subsection{Image Cropping:}
Because the methods used to detect the lattice type depend on the number of points in each image, the images needed to be cropped in such a way that nominally, the same number of centers were obtained each time \cite{Motta2018}. This was accomplished by using the speed and frequency from the PVST process to compute the expected number of pixels per circle. The highest speed was used to set an upper bound on the number of points per image, and this was held constant among the images. This information was then used to compute the pixel dimensions required to obtain a specified grid in each image and the images were cropped accordingly.

\subsection{Nominal Grid: }
The PVST process parameters that were used to generate the surface in each image were then used to plot a nominal (expected) grid on top of the actual image to visualize the difference between the nominal and the resulting grids. The nominal grid was created by first placing a datum point on top of the experimentally found center at the upper left center in the image, then computing the locations of the other points based on the nominal process parameters. An example plot of the nominal grid for the 0\% overlap ratio image is shown in Fig.~\ref{fig:120V-0_grids} as the red triangles. It is clear that the nominal grid is not aligned with the true grid as evidenced by the slight shifts in the rows which causes a change in the lattice type.

\subsection{True Grid: }
\begin{figure}[htbp]
\centering
\includegraphics[width=0.35\textwidth,height=1\textheight,keepaspectratio]{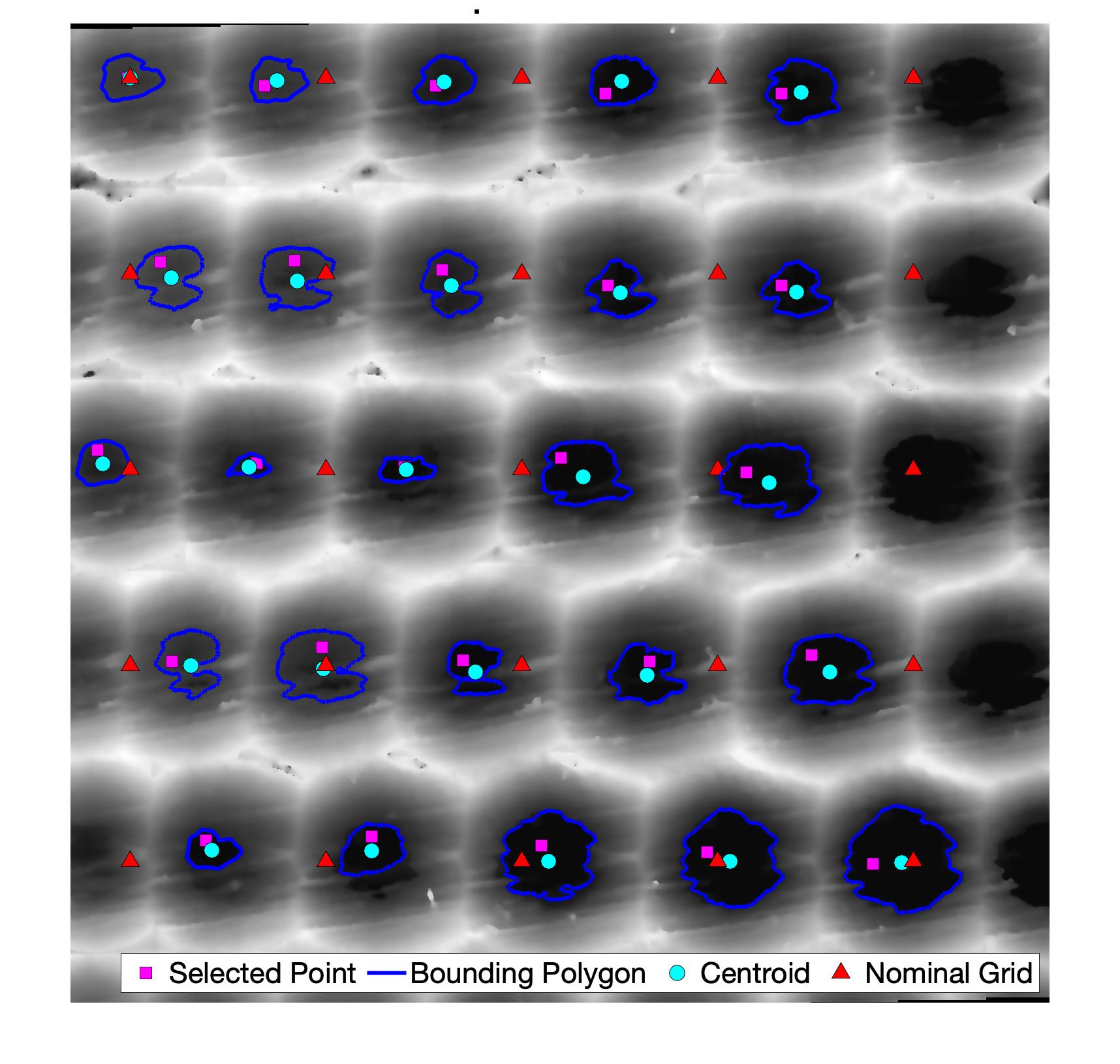}
\caption{Locating Nominal and True Lattice Centers}
\label{fig:120V-0_grids}
\end{figure}

To quantify the grid resulting from the PVST treatment, the center points of the tool indentations need to be located. This was accomplished by applying a region growing algorithm starting from a manually selected point near the center of the circle to detect a bounding polygon. The centroid of the generated bounding polygon was then used as an estimate of the center location as shown in Fig.~\ref{fig:120V-0_grids}.   The reason for manually choosing a guess for the centers' locations, instead of using the nominal locations, is shown in Fig.~\ref{fig:device}b. The figure shows that although the overlap was selected to be equal in the horizontal and the vertical direction, the columns are more widely spaced in the horizontal direction. We hypothesize that this is the result of the CNC motion whose acceleration is more smoothly varied in the direction of the scan (the vertical direction in the Fig.~\ref{fig:device}b) thus producing more precise impact locations in that direction. In contrast, we suspect that the rapid positioning motions of the CNC when moving to the next column in Fig.~\ref{fig:device}b are creating a drift in the center locations along that column that propagates every time a new column is treated which leads to incrementally shifting the whole pattern in the horizontal direction.

\section{Topological Data Analysis Based Approach}
\label{sec:TDA}
In this section, we give a brief description of persistent homology, a tool from Topological Data Analysis (TDA), specifically as it applies to point clouds since that is what we use to characterize the PVST grid. 
We refer the interested reader to more in-depth treatment of TDA in \cite{Ghrist2008, Carlsson2009b,Edelsbrunner2010, oudot2017persistence, munkres2018elements,munch2017user}.
We then summarize the information extracted from images in persistence diagrams, and we use the latter to score the PVST-treated surfaces.

\subsection{Persistent homology}
\begin{figure*}[htbp]
    \centering
    \includegraphics[width=0.85\textwidth,height=1\textheight,keepaspectratio]{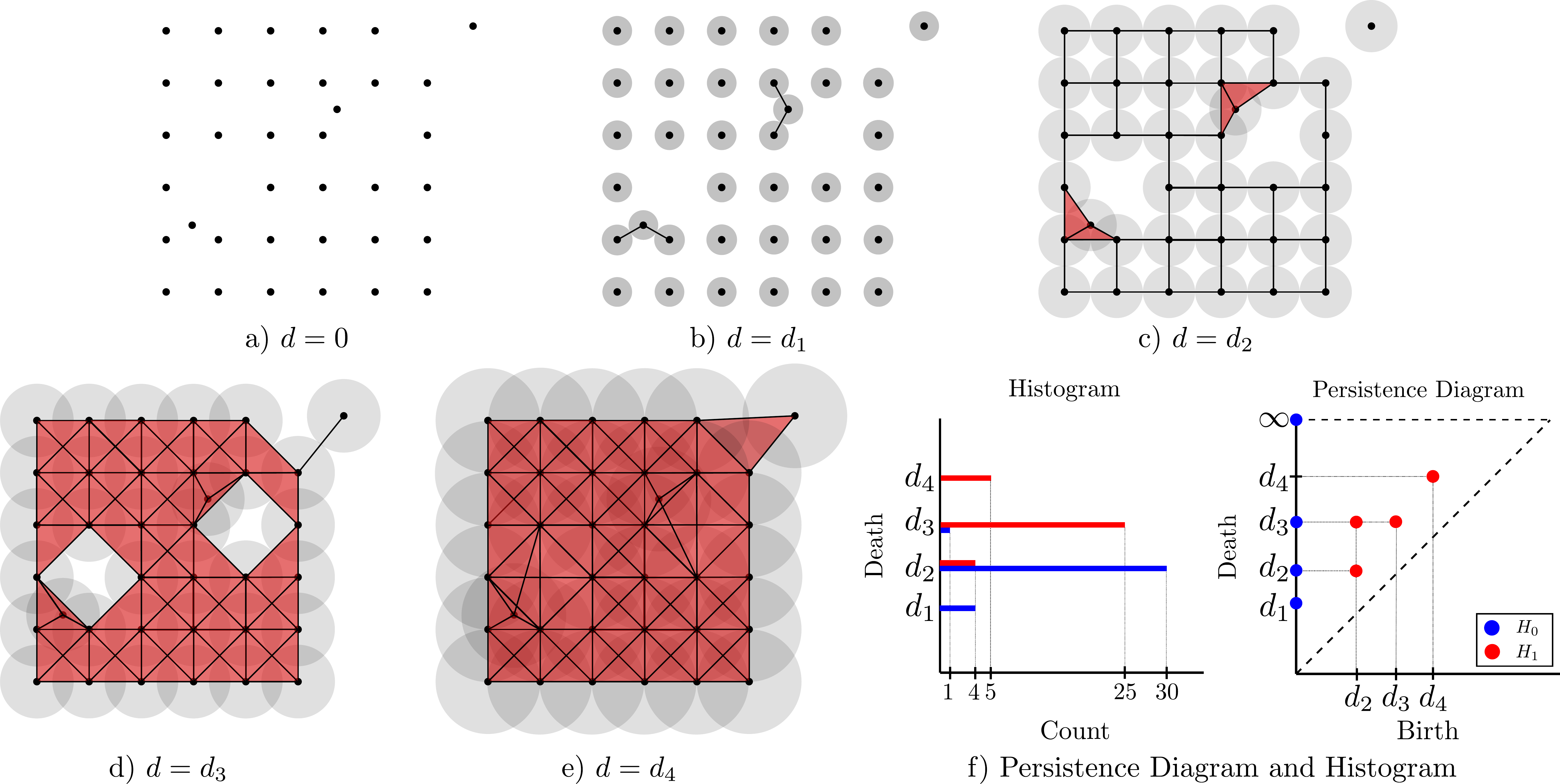}
    \caption{Persistence Example}
    \label{fig:grid_filtration}
\end{figure*}

Persistent homology, or persistence, is a tool from TDA for extracting geometric features of a point cloud such as the connectivity or the number of holes in the space as a function of a connectivity parameter. 
Specifically, consider the point cloud shown in Fig.~\ref{fig:grid_filtration}a which represent points in an almost prefect square lattice but with three perturbed points. 
Suppose that we start expanding disks of diameter $d$ around each of the points, and we monitor changes in the connectivity of the components as $d$ is increased. 

We define connectivity using the Euclidean distance in the plane between each point in the set. 
Specifically, we connect two vertices via an edge if their Euclidean distance is at least equal to the connectivity parameter $d$. 
While Fig.~\ref{fig:grid_filtration}a shows that for $d=0$ we have $36$ distinct components that emerge or are \textit{born}, Fig.~\ref{fig:grid_filtration}b demonstrates that some of these components begin to merge or \textit{die} at $d=d_1$. When $d=d_2$, the square edges connect which creates holes at the center of each square and two larger holes where the perturbations are present. Note that the third perturbation in the corner remains disconnected at this point. Once the connectivity parameter reaches $d_3$, all of the components remain connected as $d\rightarrow \infty$. 
The difference between the death value and the birth value for any component is called its \textit{the lifetime}. 

We can summarize the connectivity information using the zero-dimensional (0D) persistence diagram shown in Fig.~\ref{fig:grid_filtration}f (blue). 
The coordinates of points in this diagram are the (birth, death) values of the connectivity parameter where the number of connected components changes. This diagram shows that four of the components die when $d=d_1$ and 30 more connect at $d=d_2$ with the remaining point joining at $d_3$. These quantities are represented in the histogram shown with the persistence diagram.

Moreover, Fig.~\ref{fig:grid_filtration} shows that the grid has interesting geometric features characterized by the emergence and disappearance of holes in the plane as $d$ is varied. 
Persistence can also track the birth and death of these holes via one-dimensional (1D) persistence. 
In order to track holes, for each value of $d$ we construct a geometric object that includes the vertices themselves, pairs of points (edges) that are connected at $d$, and the 3-tuples that represent faces (geometrically, they are triangles) which get added whenever all their bounding edges are connected. 
The geometric objects constructed this way are called simplicial complexes, and varying $d$ leads to a growing sequence of them indexed by the single parameter $d$. 
For example, Fig.~\ref{fig:grid_filtration}f (red) shows that at $d=0$ only the vertices are included in the corresponding simplicial complex, and we have no holes. 
Increasing $d$ to $d_1$ in Fig.~\ref{fig:grid_filtration}b, edges are included between points whose Euclidean distance is less than or equal to $d$. 
We also fill in or include in the simplicial complex any triangles whose bounding edges all are included. 
This is shown in Fig.~\ref{fig:grid_filtration}c where four triangles are added.
At this point, when $d=d_2$, 18 loops were born that were not filled in immediately. As $d$ is increased to $d_4$, more holes emerge and are filled at the same time. This process is continued until all of the components are connected as one and no loops persist in the simplicial complex. 
The information related to the birth and death of holes is summarized in the 1D persistence diagram shown in Fig.~\ref{fig:grid_filtration}f. Notice that two of the holes born at $d_2$ are larger than the others causing them to have a longer lifetime as $d$ is varied. These larger holes are apparent in the persistence diagram with the largest vertical distance to the diagonal. 

The combination of the 0D and 1D persistence allows us to quantify the deviation of a given grid from its nominal, perfectly ordered lattice. 
For example, in a PVST process we can compute the 0D and 1D persistence of the resulting, actual grid by locating the centers of the tool on the treated surface. 
We can then compare the actual persistence values to their nominal counterparts where the latter are obtained by assuming a perfectly produced lattice with no defects or center deviations. 
This allows us to quantify the proximity of the resulting grid to the commanded grid, and gives us a tool to characterize the defects during the PVST treatment process such as misplaced or missed strikes that alter the expected pattern.

\subsection{TDA-based scores:} \label{sec:scores}
We focus on 0D ($H_{0}$) and 1D ($H_{1}$) persistence to obtain scores for texture analysis. 
In order to quantify the type of the pattern on a PVST-treated surface, a method is needed for scoring different lattices. To achieve this, point cloud persistent homology was applied to a perfect square lattice, and expressions were obtained to be used for comparison with the persistence outputs from the actual surfaces. Comparing the results from a perfect lattice to the true surface of interest allows for conclusions to be drawn about the type of lattice present due to PVST. 
To correctly compare the results from multiple images, the same number of points must be chosen in each case and the square regions containing these points must be similarly scaled, e.g., to [-1,1]$\times$[-1,1]. Otherwise, the comparisons become less meaningful because mismatched sample scaling or different number of points in each sample will strongly influence the resulting scores. 
\subsubsection{0-D Persistence}

\begin{figure*}[htbp]
    \centering
    \includegraphics[width=0.7\textwidth,height=1\textheight,keepaspectratio]{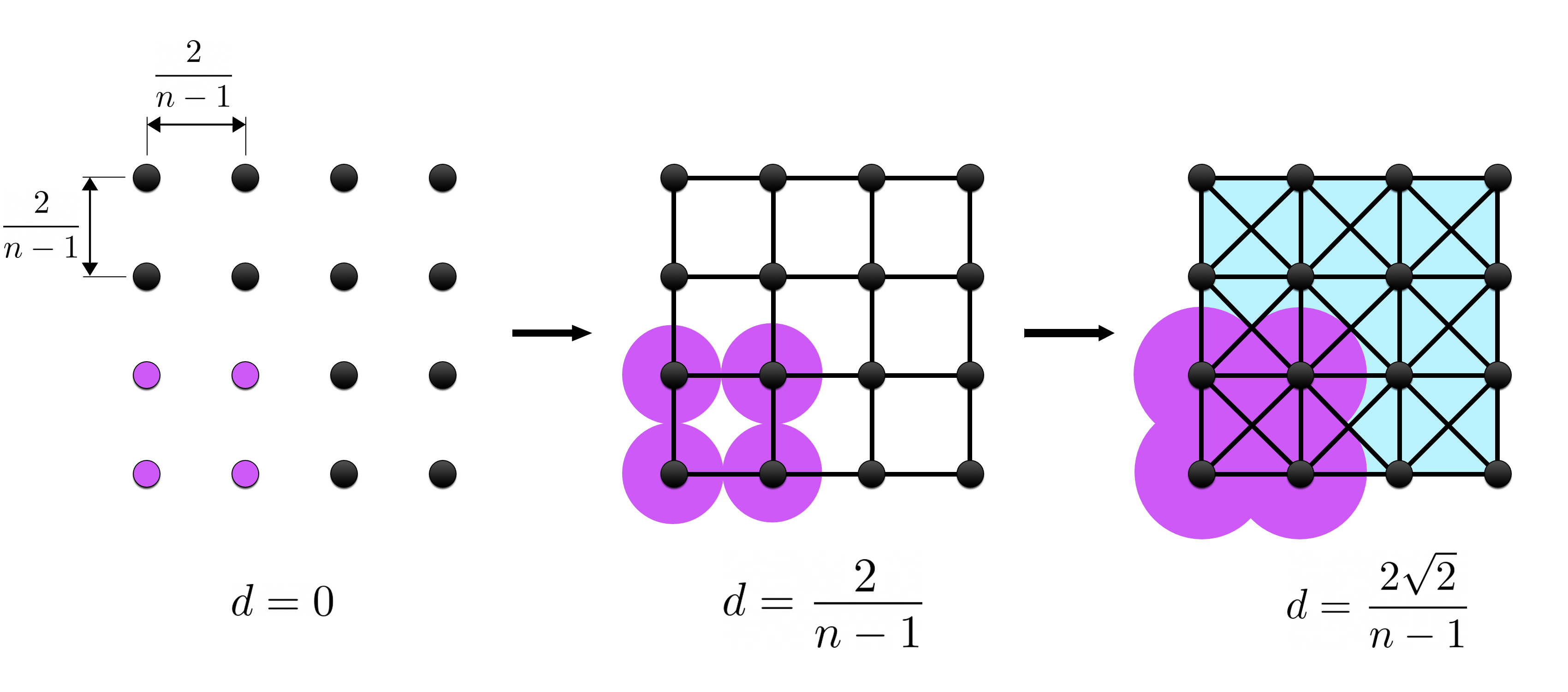}
    \caption{An illustration of the key values of the diameter $d$ at which features are born and die for a perfect Square Lattice. The expanding disks (purple) are only shown for four points to better visualize the edges and the triangles that get added at each shown $d$ value.}
    \label{fig:lattice_image}
\end{figure*}

To compute the $H_0$ persistence of a perfect square lattice, a Vietoris–Rips complex was applied to a perfect square lattice with an $n \times n$ grid of points in Fig.~\ref{fig:lattice_image} and the connectivity parameter $d$ was varied until the all of the rectangles were included in the complex. 

For the $H_0$ persistence, points were born at $d=0$ and died at $d=\frac{2}{n-1}$ where $d$ is the diameter of the expanding balls. For 0D persistence of this lattice, all of the elements are born and die at the same time. It was shown in \cite{Motta2018} that for both perfect square and perfect hexagonal lattices, the variance in the 0D persistence (lifetimes) is zero, which we express as
\begin{equation}\label{Eq:H0sum}
    Var(H_0)=0.
\end{equation}

\begin{figure}[htbp]
    \centering
    \includegraphics[width=0.45\textwidth]{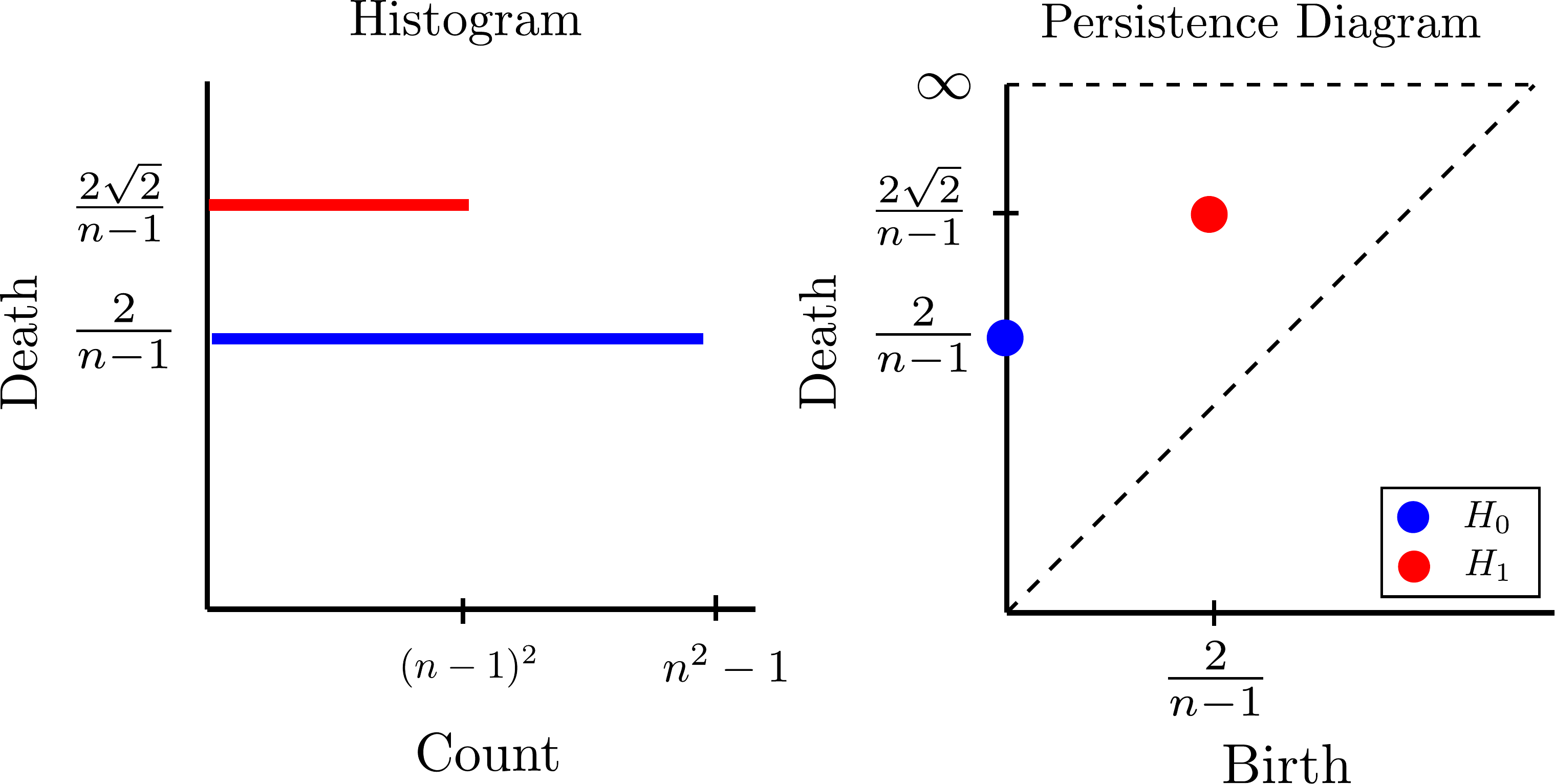}
    \caption{Theoretical Persistence Diagram for a perfect square lattice.}
    \label{fig:theoretical_persistence}
\end{figure}

This expression can be used for measuring the deviation from a square/hexagonal lattice in the presence of a non-zero variance \cite{Motta2018}. Note that the overlap ratios for the PVST images are accounted for when the nominal distance between points is computed using the in-plane speeds and frequency to locate the centroids of the circles. The theoretical persistence diagram for the 0-D persistence was generated as shown in Fig.~\ref{fig:theoretical_persistence}.

Where $n^2-1$ components are born at 0 and die at $\frac{2}{n-1}$, and one object survives for all time. The $H_0$ variance was determined to have a maximum value of $\frac{1}{4}$ over all possible lattice types \cite{Motta2018} so to normalize this measure, it was multiplied by a factor of 4. 
\subsubsection{1-D Persistence}
For the 1-D persistence in a Vietoris–Rips complex, the presence of loops is of interest. Figure~\ref{fig:lattice_image} shows that all of the loops are born at $d=\frac{2}{n-1}$ and die at $d=\frac{2\sqrt{2}}{n-1}$ giving the following lifetime for each individual loop
\begin{equation}
    H_{1}=\frac{2\sqrt{2}}{n-1}-\frac{2}{n-1}.
\end{equation}

If the grid of interest is $n \times n$ points, the total number of loops $k$ present for a perfect rectangular lattice is $k=(n-1)^2.$
The loop lifetimes provide insight into the density of the lattice in the image as a higher point density would have smaller loops. To incorporate all of the loop lifetimes into one measure, the sum of all individual lifetimes is computed. For a perfect lattice, the sum can be multiplied by the number of loops because in a perfect lattice all of the loops have the same lifetime. Therefore, Eq.~\eqref{eq:H1sum} can be used to quantify the type of lattice in the image where it is a maximum value for a square lattice and 0 for a hexagonal lattice \cite{Motta2018}.
\begin{equation}\label{eq:H1sum}
    \sum H_1 = 2(\sqrt{2}-1)(n-1).
\end{equation}

The 1-D persistence diagram for the rectangular lattice is shown in Fig.~\ref{fig:theoretical_persistence}. For the perfect lattice, $k$ loops are born at $\frac{2}{n-1}$ and die at $\frac{2\sqrt{2}}{n-1}$.
If the $H_1$ lifetimes are smaller than the perfect lattice lifetime, this would indicate the presence of shifts in the lattice type (i.e., some of the points are shifted allowing some of the loops to prematurely close). It has been shown that this measure is equal to zero for a perfect hexagonal lattice and achieves a maximum value for a square lattice~\cite{Motta2018}. In our case, the normalization factor was a function of the number of points in a row or column of the grid as shown previously. For this reason, the sum of 1D persistence lifetimes was divided by the expression shown in Eq.~\eqref{eq:H1sum}. Because the $H_1$ sum is nonzero for a square lattice, the CPH score approach presented in \cite{Motta2018} for hexagonal lattices cannot be used due to the underlying nominal lattice being square, and a square lattice cannot be detected with a single measure of order. Together, the $H_0$ and $H_1$ measures were used to provide scores for classifying the type of lattice. For example, if both scores are close to zero the lattice is mostly hexagonal. If the $H_1$ sum is close to 1 and the $H_0$ variance is small, the lattice is mostly square and if both measures are close to one the lattice is neither square nor hexagonal. Expressions for the computed scores used with the PVST centers point clouds are given by

shown in equations~\eqref{eq:norm_var} and \eqref{eq:norm_h1}.

\begin{subequations}
\begin{equation}\label{eq:norm_var}
    \overline{H_0} = 4Var(H_0),
\end{equation}
\begin{equation}\label{eq:norm_h1}
   \overline{H_1}= \frac{1}{2(\sqrt{2}-1)(n-1)}\Sigma H_1,
\end{equation}
\end{subequations}
where $\overline{H_0}$ is the normalized 0-D persistence score, and $\overline{H_1}$ is the normalized 1-D persistence score. These scores were used to detect the presence of square and hexagonal lattices in the PVST images and quantify the relative magnitudes of each type.

\section{Results}
\label{sec:Results}

Point cloud persistence was applied to the nominal and actual grids from the PVST images and the corresponding persistence diagrams and histograms were generated. The scores from Section~\ref{sec:scores} were then used with the computed statistics from the persistence output to quantify the lattice types. 

\begin{figure*}[htbp]
\centering
\begin{minipage}[t]{\textwidth}
\centering
\includegraphics[width=0.48\textwidth]{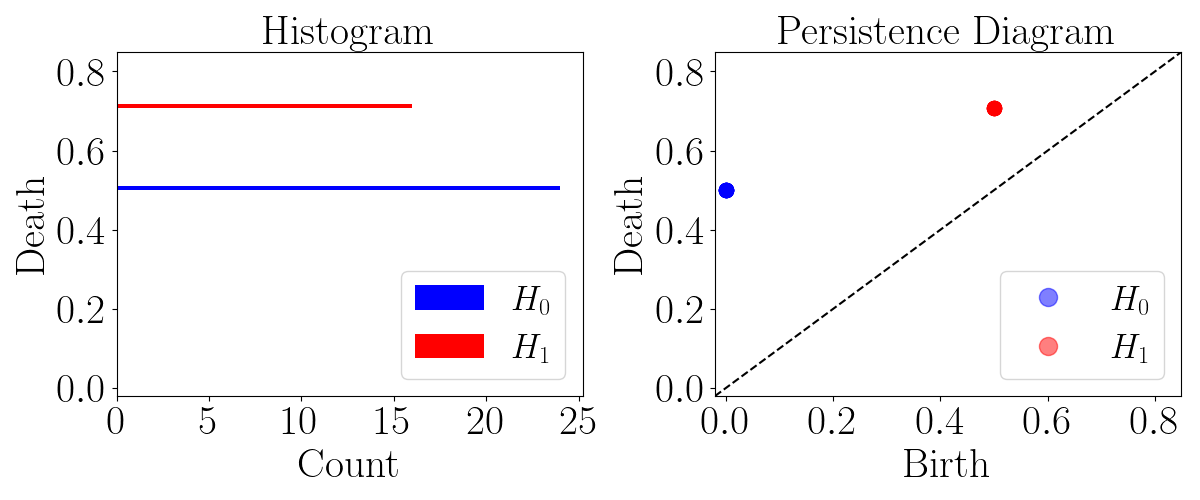}
\includegraphics[width=0.48\textwidth]{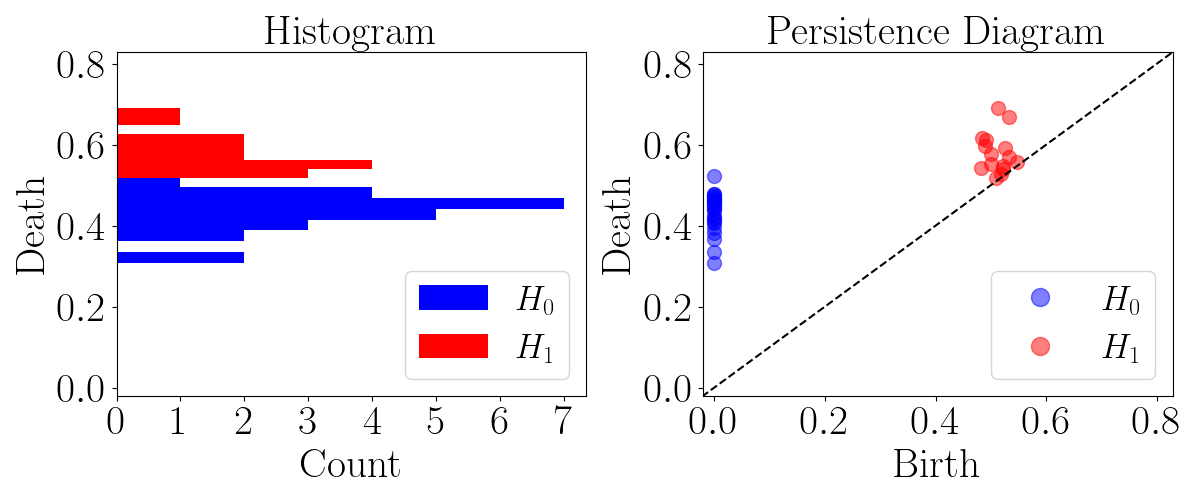}
\end{minipage}

\begin{minipage}{0.48\textwidth}
\centering
(a) Nominal Grid Persistence Diagram.
\end{minipage}
\begin{minipage}{0.48\textwidth}
\centering
(b) True Persistence Diagram 0\% Overlap.
\end{minipage}

\begin{minipage}[t]{\textwidth}
\centering
\includegraphics[width=0.48\textwidth]{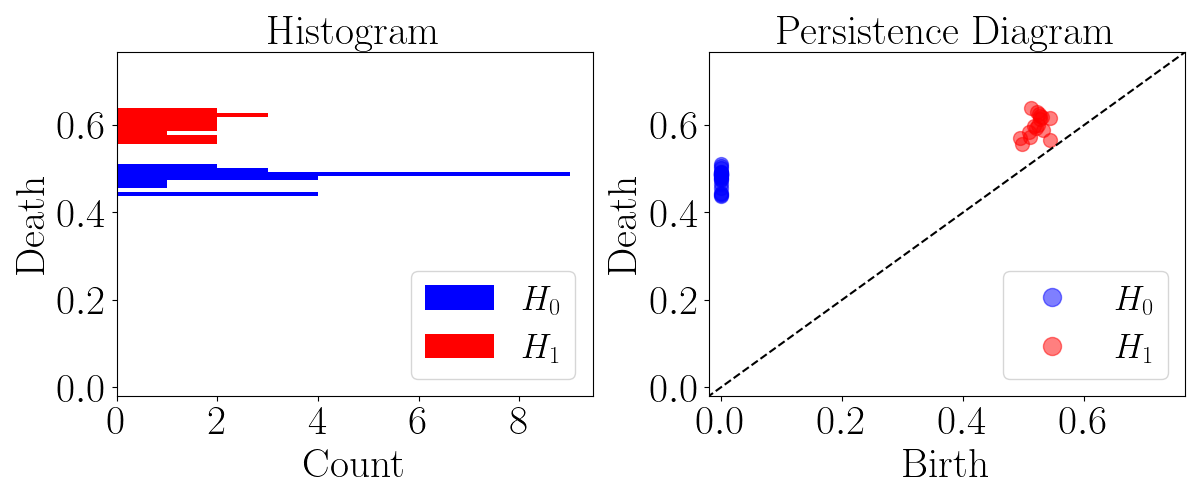} 
\includegraphics[width=0.48\textwidth]{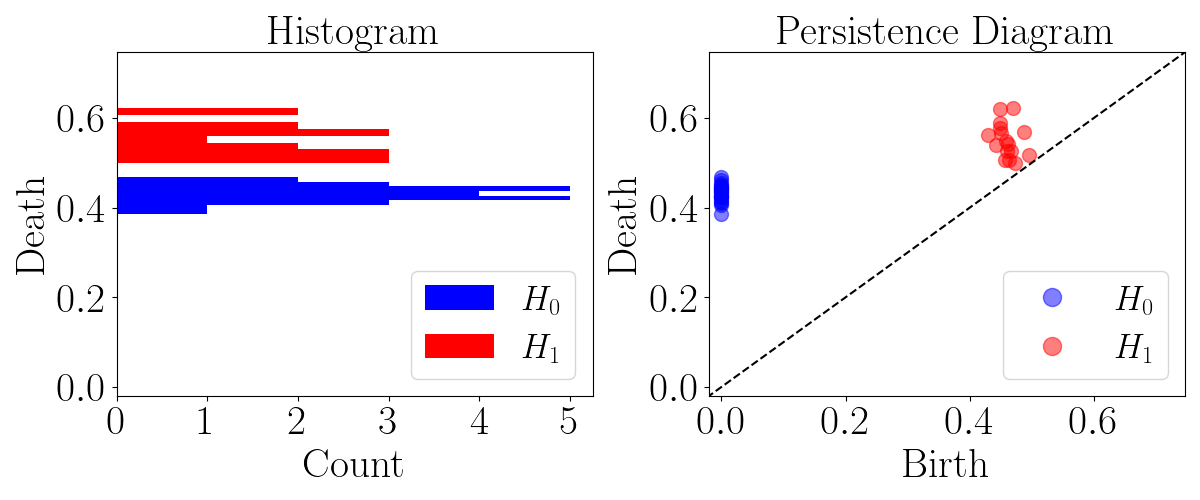}
\end{minipage}

\begin{minipage}{0.48\textwidth}
\centering
(c) True Persistence Diagram 25\% Overlap.
\end{minipage}
\begin{minipage}{0.48\textwidth}
\centering
(d) True Persistence Diagram 50\% Overlap.
\end{minipage}
\caption{The resulting persistence diagrams and the corresponding histograms for (a) a nominal grid, (b) PVST with $0$\% overlap, (c) PVST with $25$\% overlap, (d) PVST with $50$\% overlap.}
\label{fig:PDs}
\end{figure*}

\subsection{Persistence Diagrams}
The generated point clouds from Section~\ref{sec:data_processing} were passed into Ripser, a python library used to generate persistence diagrams from point cloud data with a Vietoris-Rips Complex \cite{Bauer2021Ripser}. First, nominally generated grid point clouds were passed to Ripser to verify the expressions obtained in Section~\ref{sec:scores}. The same persistence diagram resulted for all three of the grids shown in Fig.~\ref{fig:PDs}a, because the data was scaled to the same region. Histograms were plotted beside the persistence diagrams to illustrate repeated points in the diagram. We see that the 0D persistence pairs have a birth time of 0 and a death time at 0.5 which is consistent with the predicted results from Fig.~\ref{fig:lattice_image} when $n=5$. We then plotted the persistence diagrams for the actual grids obtained from the region growing algorithm. These persistence diagrams are shown in Fig.~\ref{fig:PDs}b--d. It was clear from the 0\% overlap ratio image, the point density was larger than expected which was captured by the 1D persistence diagram loops having a lower lifetime as a result of the centers being closer together in this image. The 0D persistence pairs in the 0\% overlap image appear to have a significantly larger spread when compared to the other overlap ratio results. This was likely due to some of the rows shifting into a hexagonal lattice causing some of the components to connect before they would if the lattice were square. 

\subsection{Measures of Order}
Equations~\eqref{eq:norm_var} and \eqref{eq:norm_h1} were used to compute measures of order based on the experimental persistence information generated from the point clouds. The computed measures of order for each image are shown in Table~\ref{tab:measures}. The $\overline{H_0}$ scores were all significantly smaller than 1 indicating that all of the PVST tests considered in this paper produced lattices that were somewhere between square and hexagonal. With the information from the 0D persistence score, the $\overline{H_1}$ score allows for the lattice type to be located on the lattice figures in \cite{Motta2018} showing that the 0\% overlap ratio image was the closest to a hexagonal lattice with the larger overlap images corresponding to images closer to square relative to the first image. It should be noted that all of the $\overline{H_1}$ scores were below 0.5 which meant that the resulting lattices were predominantly closer to hexagonal than square. Another way to interpret this measure is to treat it as a percentage of square lattice present in the image for $\overline{H_0}\approx 0$. With this interpretation, the images had 31.4\%, 37.9\% and 44.1\% square lattice respectively with the remaining proportion being hexagonal. 

\begin{table}[htbp]
\caption{\label{tab:measures}Measures of Order for the experimental data.}
\centering
\begin{tabular}{|c|c|c|c|} 
  \hline
  Image & 0\% Overlap & 25\% Overlap & 50\% Overlap \\ 
  \hline
  $\overline{H_0}$ & $ 2.29\mathrm{e}{-3}$ & $0.405\mathrm{e}{-3}$ & $0.337\mathrm{e}{-3}$\\ 
  \hline
  $\overline{H_1}$ & 0.314 & 0.379 & 0.441\\ 
  \hline
\end{tabular}
\end{table}

\section{Conclusion}
\label{sec:Conclusion}
Topological approaches were applied to characterizing texture of images obtained from PVST-treated surfaces to determine the underlying lattice type. Our exploratory results show that using TDA for surface texture characterization can be beneficial for quantifying the lattice shape obtained from a PVST sample. Scores used in this paper allowed for direct quantification of the proportions of different lattice types present in a PVST surface which agreed with the qualitative examination of the images. The information gained from applying this analysis removes ambiguity in determining the true lattice shape and can be used for gaining insight into process control, and gauging improvement in the regularity of the PVST process. 

The authors plan to continue work on this topic in the future to further automate the process for detecting the true PVST centers. Future work also includes expanding the analysis of PVST surfaces through quantifying the roundness of the resulting tool indentations at different level sets, and examining the consistency of the the striking depths.

\bibliographystyle{ieeetr}  
\bibliography{references}  %

\end{document}